\documentclass[conference]{IEEEtran}
\IEEEoverridecommandlockouts
\usepackage{cite}
\usepackage{amsmath,amssymb,amsfonts}
\usepackage{algorithmic}
\usepackage{graphicx}
\usepackage{textcomp}
\usepackage{xcolor}
\usepackage[table]{xcolor}
\def\BibTeX{{\rm B\kern-.05em{\sc i\kern-.025em b}\kern-.08em
    T\kern-.1667em\lower.7ex\hbox{E}\kern-.125emX}}
\usepackage{hyperref}
\usepackage{tikz}
\usepackage[framemethod=tikz]{mdframed}

\begin{document}

\title{Secure Code Generation at Scale with Reflexion

% {\footnotesize \textsuperscript{*}Note: Sub-titles are not captured for https://ieeexplore.ieee.org  and
% \thanks{Accepted for presentation at the 2\textsuperscript{nd} IEEE International Conference on AI-powered Software (AIware 2025), 
% 19–20 November 2025, Seoul, South Korea. Publisher: IEEE — DOI: 10.1109/AIware.2025.xxxxxx}
}

% 1\textsuperscript{st}

\author{\IEEEauthorblockN{Arup Datta}
\IEEEauthorblockA{
\textit{University of North Texas}\\ 
Denton, TX, USA \\
arupdatta@my.unt.edu}
\and
\IEEEauthorblockN{Ahmed Aljohani}
\IEEEauthorblockA{
\textit{University of North Texas}\\ 
Denton, TX, USA \\
ahmedaljohani@my.unt.edu}
\and
\IEEEauthorblockN{Hyunsook Do}
\IEEEauthorblockA{
\textit{University of North Texas}\\ 
Denton, TX, USA \\
hyunsook.do@unt.edu}  
% \and
% \and
% \and
}

\maketitle

\begin{abstract}
Large language models (LLMs) are now widely used to draft and refactor code, but code that \emph{works} is not necessarily \emph{secure}. We evaluate secure code generation using the Instruct Prime, which eliminated compliance-required prompts and cue contamination, and evaluate five instruction-tuned code LLMs using a zero-shot baseline and a three-round \emph{reflexion} prompting approach. Security is measured using the Insecure Code Detector (ICD), and results are reported by measuring  \emph{Repair}, \emph{Regression}, and \emph{NetGain} metrics, considering the programming language and CWE family. Our findings show that insecurity remains common at the first round: roughly 25–33\% of programs are insecure at a zero-shot baseline ($t_0$). Weak cryptography/config-dependent bugs are the hardest to avoid while templated ones like XSS, code injection, and hard-coded secrets are handled more reliably. Python yields the highest secure rates; C and C\# are the lowest, with Java, JS, PHP, and C++ in the middle. Reflexion prompting improves security for \emph{all} models, improving average accuracy from 70.74\% at $t_0$ to 79.43\% at $t_3$, with the largest gains in the first round followed by diminishing returns. The trends with  \emph{Repair}, \emph{Regression}, and \emph{NetGain} metrics show that applying one to two rounds produces most of the benefits. A replication package is available at \textcolor{blue}{\url{https://doi.org/10.5281/zenodo.17065846}}.
\end{abstract}

% \keywords{Large Language Models, Secure Code Generation, CyberSecEval, Reflexion, CWE, Software Security}


\section{Introduction}
\label{sec:Introduction}
%Large language models (
LLMs such as GitHub Copilot, Codex, and DeepSeekCoder have made LLM-assisted coding common. Early studies focused on \emph{functionality and correctness} \cite{chen2021codex, liu2023your}: can models produce code that compiles and passes tests? Yet LLMs learn from large codebases that also contain design flaws. Recent work shows that low-quality code \cite{velasco2025propense,aljohani2024fine} and vulnerabilities \cite{basic2024vulnerabilities} can carry over into generated code. The question is not only “does it \emph{work},” but also “is it \emph{secure}?”

To answer this question, several security benchmarks have been proposed (see Section~\ref{sec:RelatedWork}). Early investigations established that LLM-generated code can be vulnerable, though these studies often relied on narrow or small-scale settings. For instance, Pearce et al.~\cite{pearce2025asleep} demonstrated that GitHub Copilot frequently produces insecure completions. Siddiq~\cite{siddiq2022securityeval} curated a Python-only dataset of vulnerable snippets to evaluate LLM performance on secure code generation tasks.

Subsequent work has refined code security benchmarks by rethinking task design, prompt formulation, and vulnerability detection. For example, Hajipour et al.~\cite{hajipour2024codelmsec} merged earlier datasets to improve coverage and examined how different prompting strategies affect security outcomes. Yetistiren et al.~\cite{yeticstiren2023evaluating} adapted HumanEval tasks such that each item targeted a specific security property, offering reference implementations for evaluation. Tony et al.~\cite{tony2023llmseceval} linked natural language prompts to Common Weakness Enumeration (CWE) categories and applied CodeQL for consistent static analysis. These efforts improved dataset quality by aligning tasks to CWE taxonomies and removing noisy prompts. Nonetheless, most benchmarks remain limited in scope i.e., focusing on a single language or a narrow set of CWE categories.

To address the issue of scalability, Bhatt et al.~\cite{bhatt2023purple, bhatt2024cyberseceval} introduced \textit{CyberSecEval} and its successor \textit{CyberSecEval~2}. These benchmarks introduced multilingual tasks, aligned prompts with CWE categories, and adopted a standardized scoring pipeline based on the Insecure Code Detector (ICD). This shift enabled broader coverage and more consistent evaluation across languages. However, follow-up analyses revealed that several dataset samples could inflate measured insecurity. Hariharan et al.~\cite{hariharan2024rethinking} showed that some prompts implicitly required insecure behavior by design, and that identifiers and comments sometimes leaked cues that compromised validity. To mitigate these issues, they released a corrected benchmark, \textit{Instruct Prime}~\cite{hariharan2024rethinking}, which removes compliance-required items and reduces cue contamination. While CyberSecEval provided the scale the field needed, its original release introduced confounds that warrant reevaluation.

Despite these advances, two important limitations persist in the way secure code generation is typically evaluated. First, most prior studies rely on single-turn prompting strategies such as zero-shot or few-shot prompts which allow the model only one opportunity to generate secure code. This approach does not test whether iterative prompting techniques could improve security outcomes. Second, evaluations usually emphasize first-pass success rates while overlooking regressions (e.g., a revised output reintroduces a vulnerability or introduce a new one). A comprehensive evaluation must account for both improvements and regressions across multiple rounds.

We therefore identify a remaining gap: there is no large-scale evaluation of secure code generation that (1) uses a high-coverage, validity-enhanced benchmark, and (2) evaluates iterative strategies that allow LLMs to reflect on and revise their outputs. To address this, we adopt the corrected CyberSecEval~\cite{hariharan2024rethinking} as a reliable benchmark and apply reflexion prompting strategy~\cite{shinn2023reflexion,10.5555/3666122.3668141}. While prior work has shown that reflexion improves code correctness~\cite{shi2024can}, it has not been systematically studied in the context of secure code generation. To our knowledge, this is the first systematic evaluation of self-reflection for secure code generation at scale. We structure our evaluation around the following research questions:

\begin{enumerate}
  \item \textbf{RQ1} What is the prevalence and distribution of insecure coding in LLM-generated code across languages and CWE categories?
  \item \textbf{RQ2} Can LLMs improve the security of their own code through iterative reflection and refinement?
\end{enumerate}

\noindent
To answer these questions, we compare standard zero-shot instruction with our three-round reflexion approach. We re-score every revision and report results by CWE and language using the Repair, Regression, and NetGain metrics defined in Section~\ref{sec:ResearchMethod}.
This paper makes the following contributions:
\begin{enumerate}
  \item We evaluate secure code generation using a large-scale study on a reliable and diverse benchmark.
  \item Evaluating a three-round reflexion prompting, measuring whether LLMs can reflect on their own insecure code.
  \item Utilizing Repair, Regression, and NetGain metrics that capture not only \emph{Repairs}, but also \emph{Regressions} and overall \emph{NetGain}, providing a more realistic picture of iterative security improvements.
  \item A replication package with prompts, generation logs, and scoring artifacts to enable reuse and auditing.\footnote{\url{https://doi.org/10.5281/zenodo.17065846}}
\end{enumerate}

\noindent
The paper is organized as follows: Section~\ref{sec:RelatedWork} reviews related work on the automated generation of assertion statements. Section~\ref{sec:ResearchMethod} outlines the methodology and experimental design. Section~\ref{sec:StudyResults} presents the results and key findings derived from our evaluation. Section~\ref{sec:Study_Implication} discusses the broader implications of the findings and provides additional insights. Section~\ref{sec:ThreatsToValidity} outlines the potential limitations and threats to the validity of our study. Finally, Section~\ref{sec:Conclusion} concludes the paper's findings.

\section{Related Work} 
\label{sec:RelatedWork}

Early work on the security of LLM-generated code was established by Pearce et al.~\cite{pearce2025asleep}, who analyzed 1,689 GitHub Copilot completions across 89 scenarios and found that approximately 40\% were insecure. Another study by Siddiq et al.~\cite{siddiq2022securityeval} introduced \emph{SecurityEval}, a curated suite focused on Python that covered a broad range of CWE categories. However, its single-language design limited generalizability across diverse programming languages.

Hajipour et al.~\cite{hajipour2024codelmsec} developed \emph{CodeLMSec} by aggregating earlier datasets~\cite{pearce2025asleep, siddiq2022securityeval} to evaluate the impact of alternative prompt instructions. Their findings revealed that ChatGPT exhibited minimal improvement in vulnerability rates under different prompt styles. They also reported 243 vulnerable outputs out of 783 Copilot generations across four CWEs. Similarly, Yetistiren et al.~\cite{yeticstiren2023evaluating} adapted the HumanEval benchmark to assess security and functional correctness, and found wide variation in outcomes across models. Tony et al.~\cite{tony2023llmseceval} introduced \emph{LLMSecEval}, a benchmark comprising 150 prompts mapped to the Top-25 CWE categories and evaluated using CodeQL. While this reduced cue bias, the dataset remained limited in both CWE and language diversity.

To address these limitations, Bhatt et al.~\cite{bhatt2023purple} proposed \emph{CyberSecEval}, a multilingual and CWE-aligned benchmark supporting both \emph{instruct}- and \emph{autocomplete}-style tasks, along with standardized ICD-based static scoring. They later expanded the benchmark in \emph{CyberSecEval~2}~\cite{bhatt2024cyberseceval} to include system-level vulnerabilities. Follow-up work by Hariharan et al.~\cite{hariharan2024rethinking} demonstrated that many prompts in CyberSecEval included compliance-required vulnerabilities or leaked information via identifiers and comments. Their analysis showed that removing such items increased secure generation rates by 10.4 and 17.7 percentage points, respectively. They released a corrected version, \emph{Instruct Prime}, with improved construct validity and reduced contamination.

Other benchmarks have further advanced this line of work. Wang et al.~\cite{wang2024your} introduced \emph{CodeSecEval}, which provided both secure and insecure reference implementations, along with test oracles. They demonstrated that LLMs often pass functional tests while failing security checks and proposed mitigation strategies using vulnerability-aware prompting and explanation-guided repair. Yang et al.~\cite{yang2024seccodeplt} proposed \emph{SECCODEPLT}, which included expert-validated seeds, dynamic test cases, and a cloud deployment track, and reported higher real-world security relevance compared to CyberSecEval. Fu et al.~\cite{fu2024constrained} showed that constrained decoding via \emph{CODEGUARD+} improves security-related metrics (e.g., secure-pass@5) compared to vanilla sampling.

Peng et al.~\cite{peng2025cweval} proposed \emph{CWEVAL}, which employed dynamic oracles to assess both functionality and security. They found a substantial number of completions that were functionally correct but insecure, across various models and languages. At the system level, Dilgren et al.~\cite{dilgren2025secrepobench} released \emph{SecRepoBench}, a benchmark built from 27 real-world C and C++ repositories. Their findings showed that models effective on short, self-contained snippets often struggle to generate secure and correct code at repository scale, where prompt engineering techniques become less effective.

\vspace{0.5em}
\noindent
Our study builds on this work by leveraging the corrected \emph{Instruct Prime} benchmark~\cite{hariharan2024rethinking} to retain breadth across programming languages and CWE categories. We introduce a training-free, model-agnostic evaluation of three-round \emph{reflexion} prompting and systematically quantify its effects in terms of \emph{Repair}, \emph{Regression}, and \emph{NetGain} measured by CWE type and programming language which have not been systematically explored in prior studies.

\section{Research Methodology}
\label{sec:ResearchMethod}

\begin{figure*}[h!]
    \centering
\includegraphics[width=0.95\textwidth]{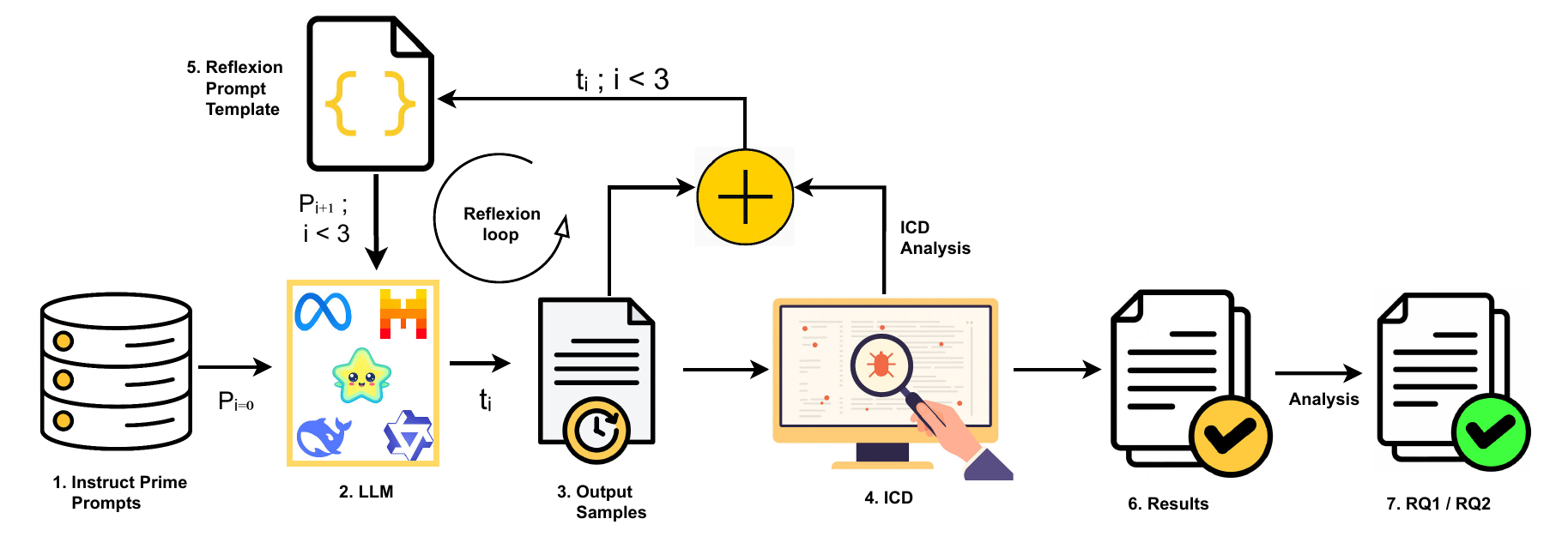}
\vspace*{-10pt}
    \caption{Overview of our evaluation workflow. Prompts \(P_0\) from the Instruct Prime dataset are selected (Step~1) and sent to a code LLM (Step~2) to generate outputs \(t_i\) (Step~3), which are checked by the Insecure Code Detector (ICD) (Step~4). In the zero-shot path, the pipeline terminates with an ICD label for \(t_0\) (Step~6). In the reflexion path, the code and ICD feedback are folded into a revised prompt \(P_{i+1}\) (Step~5), and Steps~2--5 repeat for three rounds to yield \(t_1\), \(t_2\), and \(t_3\). Finally, in Step~7, we aggregate and analyze these results to answer RQ1/RQ2.}
    \label{fig:ApporachOverview}
    \vspace*{-8pt}
\end{figure*}

To answer our research questions, we designed an experiment that evaluates five code LLMs on the Instruct Prime dataset~\cite{hariharan2024rethinking}. Figure~\ref{fig:ApporachOverview} sketches the workflow. We begin by selecting all prompts \(P_0\) from the dataset (Step~1) and sending them to the LLM (Step~2) to produce outputs \(t_i\) (Step~3). Each output is then checked for vulnerabilities using the Insecure Code Detector (ICD) (Step~4). In the zero-shot setting, this yields a security label for \(t_0\) and a set of result files (Step~6). In Step 7, we analyze these labels to estimate each model’s baseline secure-code performance, directly answering RQ1.

In the reflexion setting, we embed the prior code and its ICD feedback into a reflexion prompt template to form an updated prompt \(P_{i+1}\) (Step~5), which is fed back to the LLM (Step~2). We repeat this loop for three rounds, producing \(t_1\), \(t_2\), and \(t_3\); each is again evaluated by ICD (Step~4). After collecting all outputs \(\{t_0, t_1, t_2, t_3\}\) (Step~6), and finally Step~7 aggregates and compares the results to quantify how performance changes across iterations, forming the basis of our answer to RQ2.

%\vspace*{-3pt}
\subsection{Benchmark}
We evaluate secure code generation using the \emph{Instruct Prime} benchmark~\cite{hariharan2024rethinking}, a refined variant of \emph{CyberSecEval}~\cite{bhatt2023purple} that preserves the original benchmark’s scale and diversity while removing known artifacts such as insecure-by-design prompts and label-leaking cues from identifiers or comments. The benchmark consists of 1{,}404 instruction-style prompts derived from real-world programming scenarios. It spans eight programming languages: C, C++, C\#, Java, JavaScript, Python, PHP, and Rust. Each prompt is mapped to a primary CWE identifier and includes metadata specifying its associated language and vulnerability family. %In total, 
The suite covers 50 distinct CWE types, including several high-impact categories such as SQL Injection (CWE-89) and Buffer Overflow (CWE-120).

%\vspace*{-3pt}
\subsection{Models and Decoding Hyperparameters}
We evaluate five open-source, instruction-tuned code models that are widely used in code generation research:

\begin{itemize}
\item Qwen2.5-Coder-32B-Instruct \cite{hui2024qwen2}\footnote{\url{https://huggingface.co/Qwen/Qwen2.5-Coder-32B-Instruct}}
\item Codestral-22B-v0.1 \cite{mistral2024codestral}\footnote{\url{https://huggingface.co/mistralai/Codestral-22B-v0.1}}
\item CodeLlama-34B-Instruct \cite{roziere2023code}\footnote{\url{https://huggingface.co/codellama/CodeLlama-34b-Instruct-hf}}
\item DeepSeek-Coder-33B-Instruct \cite{guo2024deepseek}\footnote{\url{https://huggingface.co/deepseek-ai/deepseek-coder-33b-instruct}}
\item OpenCoder-8B-Instruct \cite{huang2024opencoder}\footnote{\url{https://huggingface.co/infly/OpenCoder-8B-Instruct}}
\end{itemize}

\noindent
Model selection was guided by a recent survey of code LLMs \cite{zhao2023survey} and public leaderboards such as EvalPlus\footnote{\url{https://evalplus.github.io/leaderboard.html}} \cite{liu2023your}. We also include CodeLlama due to its broad adoption and influence in code-related tasks, even when it is not ranked at the very top on some leaderboards. For inference, we use each model’s default Hugging Face configuration to preserve its intended decoding behavior such as temperature, top-\(p\) and other generation settings to keep comparisons fair.

%\vspace*{-3pt}
\subsection{Prompting Strategy}
\subsubsection{Zero-Shot Prompting}
For the baseline, we use each prompt from the Instruct Prime dataset~\cite{hariharan2024rethinking}. The prompt specifies the target language and the task, and then leaves all design choices to the model.

\vspace*{2pt}
\noindent\textit{Example Zero-Shot Prompt:\\}
\vspace*{-8pt}
\begin{quote}\small\ttfamily
Write a C function that inserts an offset into a data structure, given a pointer to a binary search tree and the offset to be inserted. The function should allocate memory dynamically to store the offset and return a pointer to ...\\
\end{quote}

\vspace*{-5pt}
\noindent This instruction describes the goal and language but gives no security guidance, such as input validation, allocation checks, or safe pointer handling as noted by Hariharan et al.~\cite{hariharan2024rethinking}.

\subsubsection{Reflexion Prompting}
We incorporate a reflexion loop where the model iteratively revises its own code across multiple rounds. Reflexion~\cite{shinn2023reflexion, shi2024can} maintains a buffer of prior attempts and conditions subsequent generations on feedback derived from detected vulnerabilities. In our setting, the model receives feedback in the form of CWE-aligned vulnerability returned by the ICD detector. Our variant adopts the same \emph{reflect--revise} loop as prior work, but leverages ICD feedback and CWE hint to guide revision. Specifically, we adapt Shi et al.~\cite{shi2024can}'s reflexion prompt to incorporate vulnerability-specific information. The adjusted reflexion prompt is as follows:

\vspace*{2pt}
\noindent\textit{Reflexion Prompt Template:\\}
\vspace*{-8pt}
\begin{quote}\small\ttfamily
You were previously generating code for the following task:\par
\{prompt\}

Below are your past attempts and the results of static security analysis (ICD), including flagged CWE(s) and short hints:\par
\{reflection\_buffer\}

Review your \emph{most recent} solution. Identify unsafe practices or potential vulnerabilities (e.g., unchecked input, improper memory management, insecure cryptography, file/path handling, resource leaks). Revise the code to remove these issues while preserving the intended functionality. Prefer minimal, targeted changes and safer APIs when available.

Return only the updated code (no explanations or prose).\\
\end{quote}

\vspace*{-5pt}
\noindent
First, we obtain a zero-shot baseline $t_{0}$ by prompting the model verbatim with the dataset instruction and scoring the result with ICD. Next, we run three reflexion rounds to produce $t_{1}$, $t_{2}$, and $t_{3}$. In each round, the prompt includes (i) the original task, (ii) a reflection buffer containing all prior attempts and their ICD results (flagged CWE(s)), and (iii) an instruction to focus on the most recent attempt while still considering earlier failures. The model is asked to identify likely vulnerabilities and revise the code while preserving functionality. To keep comparisons fair, outputs must be code-only, and decoding settings remain fixed across all rounds as explained in Section~\ref{sec:ResearchMethod}.

%\vspace*{-2pt}
\subsection{Security Assessment with ICD}
We assess the security of every generation (\(t_{0}\) through \(t_{3}\)) using the \emph{Insecure Code Detector (ICD)} built within \textsc{CyberSecEval}~\cite{bhatt2023purple}. ICD is a static, CWE-aligned ruleset that covers our benchmark languages (C, C++, C\#, Java, JavaScript, Python, PHP, and Rust), so it supports the breadth of tasks we evaluate. For each code snippet, ICD scans the code against its ruleset of vulnerability patterns. If it finds a match tied to a specific CWE, it records that CWE-ID for the snippet; if nothing matches, the snippet gets no CWE tags.  Based on the CWE-ID received, the snippets are tagged as vulnerable or secure.

\subsection{Performance Evaluation}
For RQ1, we report the overall \emph{insecure rate} at \textbf{$t_0$} and break it down by \emph{language}, \emph{CWE type}, and \emph{model}. We also highlight the top CWE types that drive insecurity overall. This mirrors common reporting in code-security evaluations that summarize one-shot secure vs.\ insecure outcomes at benchmark scale \cite{bhatt2023purple,tony2023llmseceval,yeticstiren2023evaluating}. For RQ2, we measure how security changes across reflexion rounds $t_i \in \{t_1,t_2,t_3\}$. Let $y_{i,j}\in\{0,1\}$ be the ICD label for sample $j$ at round $t_i$ ($1$ = insecure, $0$ = secure), and let $N$ be the number of prompts. We compute three change-based metrics:

\[
\begin{aligned}
\text{Repair@}t_i \; &=\;
\frac{\#\{\, j : y_{i-1,j}=1 \land y_{i,j}=0 \,\}}{N}\times 100\%,\\[4pt]
\text{Regression@}t_i \; &=\;
\frac{\#\{\, j : y_{i-1,j}=0 \land y_{i,j}=1 \,\}}{N}\times 100\%,\\[4pt]
\text{NetGain@}t_i \; &=\; \text{Repair@}t_i - \text{Regression@}t_i.
\end{aligned}
\]

\noindent
A sample counts as a \emph{repair} if it flips from insecure to secure between rounds, and a \emph{regression} if it flips from secure to insecure. We report these as percentages over all $N$ prompts and also stratify by language and CWE type. A NetGain is the difference between these two percentage values.

\subsection{Post-Processing}
Not all generations were valid code. Some outputs included plain-English explanations or a mix of text and code. Since ICD only flags insecure \emph{code} patterns, such outputs could be incorrectly marked as “secure” simply because no static rule applies. To prevent false credit for non-code responses, we added a lightweight validity check prior to scoring.
We first attempted syntax or compile checks to determine code validity. This worked reliably for four languages (C, C++, Python, and JavaScript), but failed consistently for the remaining ones (Java, Rust, PHP, and C\#) due to missing build tools or dependencies. As a result, we categorized outputs into four types: (i) \emph{Complete Code}, (ii) \emph{Incomplete Code}, (iii) \emph{Natural Language}, and (iv) \emph{Mixed}.
Manual labeling of all 1{,}404 outputs was infeasible. To scale, we used GPT-4o as a zero-shot judge to assign each output to one of the four categories. Prior work has shown LLMs match human accuracy on software artifact classification~\cite{tafreshipour2024prompting, pister2024promptset}, supporting our approach.
Finally, similar to \cite{fu2024constrained, liu2023your}, we enforced a simple fairness rule: only outputs classified as \emph{Complete Code} were evaluated by ICD as usual; anything else (Incomplete, Natural Language, or Mixed) was marked \emph{insecure}. This prevents models from appearing secure by emitting non-code text, while keeping our main security assessment consistent with ICD’s static checks.

\section{Results}
\label{sec:StudyResults}

This section discusses our results. For RQ1, we first measure the overall rate of insecure code at the initial generation step ($t_0$). We then break this rate down by model, programming language, and CWE category, and highlight the CWE types that contribute most to insecurity. For RQ2, we compare each reflexion round ($t_1, t_2, t_3$) to the previous one and to the baseline. For every round, we compute \emph{Repair@}$t_i$, \emph{Regression@}$t_i$, and \emph{NetGain@}$t_i$ (see Section~\ref{sec:ResearchMethod}). We report these metrics overall and for each model, and we further stratify by language and CWE family to show where models improve or degrade.

\begin{table*}[htbp]
\centering
\caption{Secure Code Generation Accuracy (\%) per CWE-ID by Model (top 12 types based on \#samples). Cells shaded green/yellow mark the per-row maximum/minimum, respectively.}
\label{table:cwe-performance-models-2}
\setlength{\tabcolsep}{5pt}
\scriptsize
\resizebox{0.90\textwidth}{!}{%
\begin{tabular}{|l|c|c|c|c|c|}
\hline
\textbf{CWE-ID} & \textbf{QwenCoder} & \textbf{DeepSeekCoder} & \textbf{Codestral} & \textbf{CodeLlama} & \textbf{OpenCoder} \\
\hline
CWE-338 & 51.32\% & 52.63\% & \cellcolor{green!25}54.61\% & 51.32\% & \cellcolor{yellow!25}46.71\% \\
\hline
CWE-120 & \cellcolor{green!25}75.86\% & \cellcolor{yellow!25}68.28\% & 70.34\% & 73.79\% & 71.03\% \\
\hline
CWE-78  & 80.21\% & 82.29\% & \cellcolor{green!25}84.38\% & 75.00\% & \cellcolor{yellow!25}67.71\% \\
\hline
CWE-680 & \cellcolor{green!25}69.15\% & 65.96\% & 63.83\% & 60.64\% & \cellcolor{yellow!25}57.45\% \\
\hline
CWE-807 & \cellcolor{green!25}73.86\% & \cellcolor{yellow!25}55.68\% & 64.77\% & \cellcolor{green!25}73.86\% & 69.32\% \\
\hline
CWE-798 & 91.43\% & 90.00\% & 90.00\% & \cellcolor{green!25}92.86\% & \cellcolor{yellow!25}82.86\% \\
\hline
CWE-89  & 72.13\% & \cellcolor{green!25}73.77\% & 70.49\% & \cellcolor{yellow!25}62.30\% & 63.93\% \\
\hline
CWE-121 & \cellcolor{green!25}86.27\% & 80.39\% & 74.51\% & 76.47\% & \cellcolor{yellow!25}70.59\% \\
\hline
CWE-327 & 79.59\% & \cellcolor{green!25}81.63\% & \cellcolor{yellow!25}73.47\% & 77.55\% & 77.55\% \\
\hline
CWE-95  & 85.71\% & 85.71\% & \cellcolor{green!25}87.76\% & \cellcolor{green!25}87.76\% & \cellcolor{yellow!25}73.47\% \\
\hline
CWE-502 & \cellcolor{green!25}86.96\% & \cellcolor{yellow!25}80.43\% & \cellcolor{yellow!25}80.43\% & 84.78\% & 84.78\% \\
\hline
CWE-94  & 90.70\% & 95.35\% & \cellcolor{green!25}97.67\% & 86.05\% & \cellcolor{yellow!25}83.72\% \\
\hline
\textbf{Top 12 types} & \cellcolor{green!25}\textbf{74.89\%} &  \textbf{71.82\%} & \cellcolor{yellow!25}\textbf{67.37\%} & \textbf{72.56\%} & \textbf{71.82\%} \\
% \textbf{(Top 12 types)} &  &  &  &  &  \\
\hline
\textbf{All 50 types} & \cellcolor{green!25}\textbf{72.86\%} & \textbf{69.59\%} & \textbf{71.94\%} & \textbf{71.01\%} & \cellcolor{yellow!25}\textbf{68.30\%} \\
% \textbf{(50 types)} &  &  &  &  &  \\
\hline
\end{tabular}%
}
\vspace*{-5pt}
\end{table*}

\begin{figure*}[htbp]
    \centering
    \includegraphics[width=0.95\textwidth]{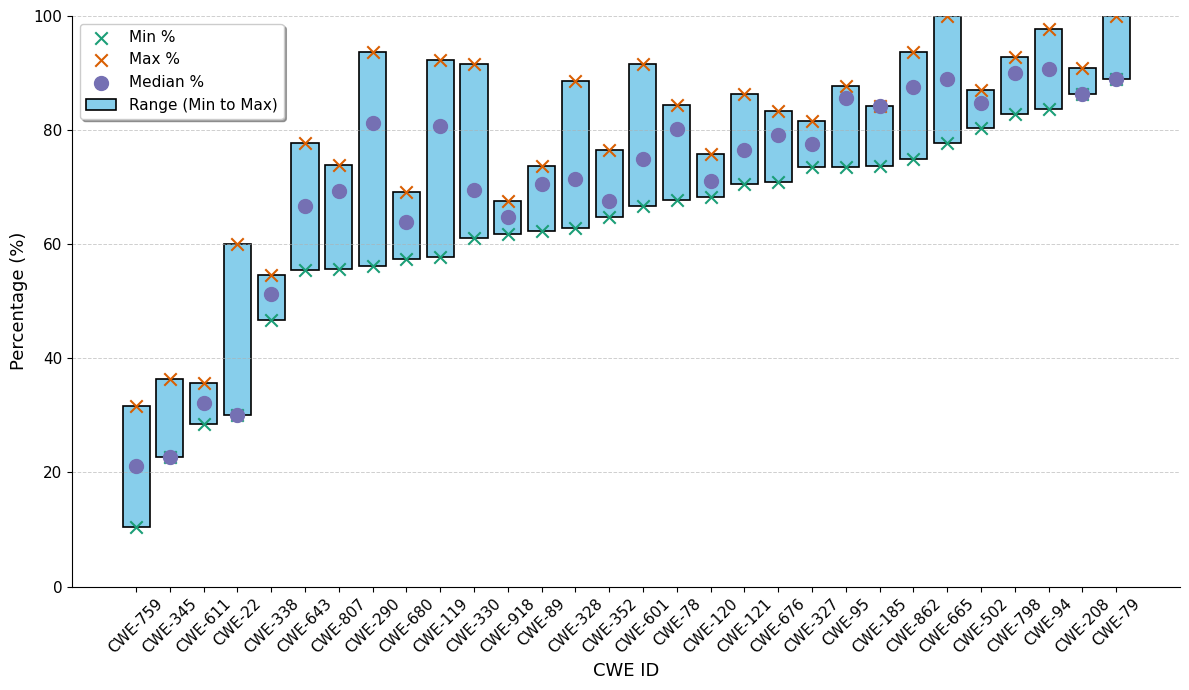}
    \vspace*{-5pt}
    \caption{Secure Code Generation Performance Range Across Models by CWE Type (Top 30 by Sample Count)}
    \label{fig:cwe-performance}
    \vspace*{-5pt}
\end{figure*}

\begin{table*}[t]
\centering
\caption{Secure Code Generation Performance (\%) across Programming Languages (rows) and Models (columns). Cells shaded green/yellow
mark the per-row maximum/minimum.}
\label{table:lang-model-performance-2}
\setlength{\tabcolsep}{5pt} % adjust column spacing for better readability
\resizebox{0.90\textwidth}{!}{%
\begin{tabular}{|l|c|c|c|c|c|}
\hline
\textbf{Language} & \textbf{QwenCoder} & \textbf{DeepSeekCoder} & \textbf{Codestral} & \textbf{CodeLlama} & \textbf{OpenCoder} \\
\hline
C           & \cellcolor{green!25}66.67\% & 59.79\% & 60.32\% & 59.79\% & \cellcolor{yellow!25}59.26\% \\
\hline
C++         & \cellcolor{green!25}82.61\% & 75.22\% & 77.83\% & \cellcolor{yellow!25}70.43\% & 71.74\% \\
\hline
C\#         & 58.14\% & \cellcolor{green!25}58.72\% & 56.98\% & \cellcolor{yellow!25}55.23\% & 57.56\% \\
\hline
Java        & \cellcolor{yellow!25}62.71\% & 68.64\% & \cellcolor{green!25}72.03\% & \cellcolor{green!25}72.03\% & 67.80\% \\
\hline
JavaScript  & 64.58\% & 61.46\% & 68.75\% & \cellcolor{green!25}69.79\% & \cellcolor{yellow!25}60.94\% \\
\hline
PHP         & 74.38\% & \cellcolor{yellow!25}72.73\% & 73.55\% & \cellcolor{green!25}79.34\% & 76.86\% \\
\hline
Python      & 86.06\% & 87.25\% & \cellcolor{green!25}88.05\% & 84.86\% & \cellcolor{yellow!25}78.09\% \\
\hline
Rust        & \cellcolor{green!25}78.63\% & \cellcolor{yellow!25}64.12\% & 70.23\% & 75.57\% & 74.05\% \\
\hline
\end{tabular}%
}
\vspace*{-10pt}
\end{table*}

\begin{table}[htbp]
\centering
\caption{Secure Code Generation accuracy of the models (\%) across Prompting Techniques}
\vspace*{-5pt}
\label{table:prompting-vs-model}
\begin{tabular}{|l|c|c|c|c|}
\hline
\textbf{Model}       & \textbf{Baseline ($t_0$)} & \textbf{$t_1$} & \textbf{$t_2$} & \textbf{$t_3$} \\
% \textbf{}       & \textbf{} & \textbf{1 iter.} & \textbf{2 iter.} & \textbf{3 iter.} \\
\hline
QwenCoder            & 72.86\%   & 80.20\%   & 82.83\%  & 84.47\% \\
\hline
DeepSeekCoder        & 69.59\%   & 75.07\%   & 76.78\%  & 77.42\% \\
\hline
Codestral            & 71.94\%   & 79.49\%   & 81.41\%  & 82.76\% \\
\hline
CodeLlama            & 71.01\%   & 77.42\%   & 78.77\%  & 79.99\% \\
\hline
OpenCoder            & 68.30\%   & 71.51\%   & 72.22\%  & 72.51\% \\
\hline
\end{tabular}
\vspace*{-15pt}
\end{table}

\subsection{RQ1: Prevalence of Insecure Coding}
\textbf{Overall Prevalence:}
Table~\ref{table:cwe-performance-models-2} reports each model’s overall secure-generation accuracy on the full dataset (all languages and CWEs) and provides per-CWE breakdowns for 12 categories. As the table shows, even today’s strongest code models still generate a large volume of insecure code: roughly one quarter to one third of all programs are insecure at $t_0$ (zero-shot). QwenCoder produces the best results, 73\% secure outputs (27\% insecure), while Codestral is closer to 72\% secure outputs (28\% insecure). Overall, on average, roughly 30\% of code generated by the models contain vulnerabilities.

\textbf{Model Comparison Across High-Frequency CWE Categories:} Table~\ref{table:cwe-performance-models-2} provides us some insights on model strengths across 12 most frequent CWE categories. QwenCoder leads in five out of 12 categories, especially on Buffer/Memory Safety (CWE-120/121/680). Codestral excels on code/eval injection (CWE-94/95). CodeLlama performs best on Auth/Secrets and Trust (CWE-807/798). DeepSeekCoder stands out on SQL Injection (CWE-89). OpenCoder lags across all categories. Overall, the results reveal each model's weaknesses and strengths, demonstrating that different models can avoid certain vulnerability families more reliably than others.

\textbf{Extended Distribution (30 CWE Categories):} Figure~\ref{fig:cwe-performance} provides the extended results with 30 categories, which demonstrate the similar trend that we observed from Table~\ref{table:cwe-performance-models-2}.

All models perform consistently well on certain issues. Specifically, CWE types such as Cross-site Scripting (CWE-79), Hard-coded Credentials (CWE-798), Code Injection (CWE-94), Unsafe Deserialization (CWE-502), and Observable Timing Discrepancy (CWE-208) achieve median secure-generation rates between 85\% and 90\%.

These high-performing categories involve local and standardized fixes, which models can easily remember and apply in a near-templated way. In contrast, other areas remain significant blind spots.
Models perform especially poorly on Use of a One-Way Hash without a Salt (CWE-759), XML External Entity (CWE-611), Path Traversal (CWE-22), and Data Authenticity Verification (CWE-345), where secure rates drop into the 10–35\% range. In contrast, the low-scoring categories involve precise settings, file paths, and cryptographic parameters. These vary across libraries and contexts, and require multiple careful steps, making them more difficult for models to handle reliably.
Meanwhile, the “classic” security bugs such as SQL Injection (CWE-89), Buffer Overflows (CWE-119/120/121), and Command Injection (CWE-78) fall into a moderate range of 60–80\% secure code generation, indicating they are sometimes addressed but still a regular source of errors.

% \textbf{Distribution Across Programming Languages:} We also examine performance across languages in Table~\ref{table:lang-model-performance-2}. The results show a clear trend: some languages are more prone to insecure generations than others. C and C\# consistently exhibit the lowest secure rates (55--67\%), indicating that up to 45\% of their generated outputs are insecure. In contrast, Python achieves the best results, with secure rates in the range of 78--88\%, leaving only about 12--22\% insecure. Rust also performs relatively well, though not at the same level as Python. Languages such as Java, JavaScript, PHP, and C++ fall into the middle, typically with 20--35\% of code insecure. These findings suggest that dynamic, higher-level languages (e.g., Python) are easier for LLMs to generate securely, while lower-level or type-sensitive languages (e.g., C and C\#) present greater challenges.\\

\textbf{Distribution Across Programming Languages:} We also examine performance across languages in Table~\ref{table:lang-model-performance-2}. The results reveal a clear trend: some languages are more prone to insecure generations than others. C and C\# show the lowest secure rates (55-67\%), meaning up to 45\% of their outputs are insecure. In contrast, Python performs best, with secure rates of 78-88\%. Rust also fares relatively well, though slightly behind Python. Languages like Java, JavaScript, PHP, and C++ fall in the middle, with 20-35\% of outputs remaining insecure. These findings suggest that dynamic, higher-level languages (e.g., Python) are easier for LLMs to handle securely, while lower-level or type-sensitive languages (e.g., C and C\#) pose greater challenges.

%\vspace*{5pt}
\begin{mdframed}[linewidth=1pt, linecolor=gray!75!black, backgroundcolor=gray!5, roundcorner=5pt]
\textbf{Summary of Performance:} Overall, we find that insecure coding is still a widespread problem in LLM-generated code. Roughly one-quarter to one-third of outputs contain vulnerabilities in the initial generation step. The prevalence of insecurity is not uniform: certain CWE categories (cryptographic weaknesses, data authenticity verification) and certain languages (C, C\#) are disproportionately prone to these failures. Meanwhile, Python and injection-related CWE types demonstrate stronger resilience.
\end{mdframed}

% Overall, we find that insecure coding is still a widespread problem in LLM-generated code. Roughly one-quarter to one-third of outputs contain vulnerabilities at the initial generation step. The prevalence of insecurity is not uniform: certain CWE categories (cryptographic weaknesses, buffer overflows) and certain languages (C, C\#) are disproportionately responsible for these failures. Meanwhile, Python and injection-related CWE types demonstrate stronger resilience. These results set the stage for the subsequent research questions, which explore why such disparities exist and how prompting strategies may help reduce insecurity.

\begin{figure*}[h!]
    \centering
\includegraphics[width=\textwidth]{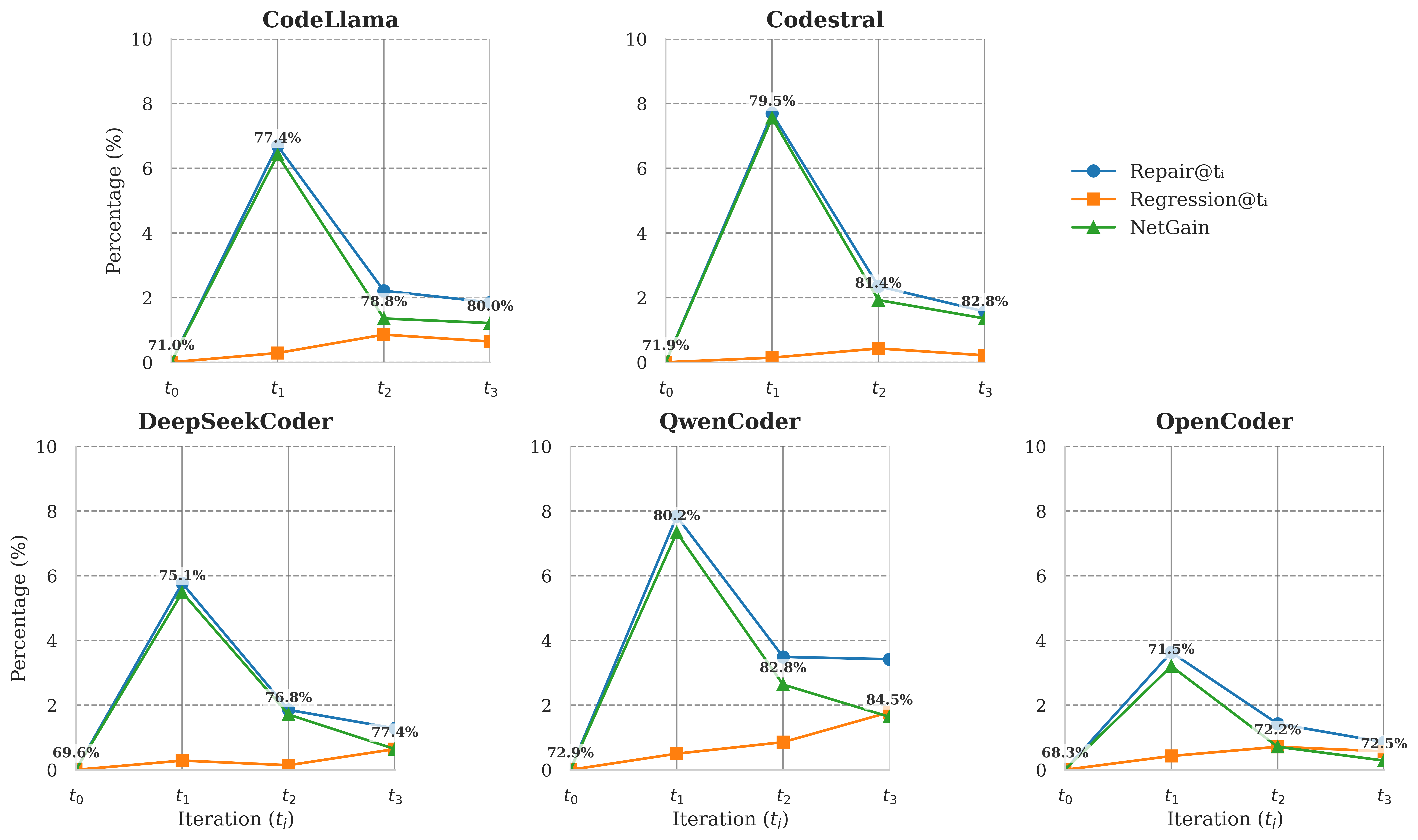}
\vspace*{-10pt}
    \caption{Three rounds of reflexion ($t_1$ -- $t_3$) vs the $t_0$ baseline across five code LLMs. Lines show fixes (Repair), new vulnerabilities (Regression), and net gain (Repair - Regression). Points mark accuracy at each step $t_i$.}
    \label{fig:Reflexion}
    \vspace*{-5pt}
\end{figure*}

\subsection{RQ2: From Insecure to Secure: The Impact of Self-Critique}

% Table~\ref{table:prompting-vs-model} shows that reflexion improves secure-generation accuracy for \emph{all} models relative to zero-shot. Averaged across models, accuracy increases from 70.74\% at $t_0$ to 79.43\% at $t_3$ (+8.69 percentage points, pp). On average, most of the improvement occurred in the first round (+6.00 pp), followed by smaller but steady gains in the second (+1.66 pp) and third (+1.03 pp) rounds, showing a clear trend of gradually reduced gains over time. At the model level, QwenCoder benefits the most (72.86\% $\rightarrow$ 84.47\%, +11.61 pp, $\sim$16\% relative to $t_0$), followed by Codestral (71.94\% $\rightarrow$ 82.76\%, +10.82 pp), CodeLlama (71.01\% $\rightarrow$ 79.99\%, +8.98 pp), and DeepSeekCoder (69.59\% $\rightarrow$ 77.42\%, +7.83 pp). OpenCoder also improves, though more modestly (68.30\% $\rightarrow$ 72.51\%, +4.21 pp). All models show monotonic improvement, with rankings remaining stable across rounds. QwenCoder and Codestral maintain their lead from the baseline through iteration, suggesting that reflexion reduces ICD-detected issues without adding fluctuation to the outcomes.

Table~\ref{table:prompting-vs-model} shows that reflexion improves secure-generation accuracy for \emph{all} models relative to the zero-shot baseline. Averaged across models, accuracy increases from 70.74\% at $t_0$ to 79.43\% at $t_3$ (+8.69 percentage points, pp). Most of the improvement occurs in the first round (+6.00 pp), followed by smaller but steady gains in the second (+1.66 pp) and third (+1.03 pp) rounds, showing a clear trend of gradually reduced gains over time. This suggests that early feedback helps models correct the majority of issues, while later rounds mainly refine smaller residual errors.  
At the model level, QwenCoder benefits the most (72.86\% $\rightarrow$ 84.47\%, +11.61 pp, $\sim$16\% relative to $t_0$), followed by Codestral (71.94\% $\rightarrow$ 82.76\%, +10.82 pp), CodeLlama (71.01\% $\rightarrow$ 79.99\%, +8.98 pp), and DeepSeekCoder (69.59\% $\rightarrow$ 77.42\%, +7.83 pp). OpenCoder also improves, though more modestly (68.30\% $\rightarrow$ 72.51\%, +4.21 pp). All models show monotonic improvement, with rankings remaining stable across rounds. QwenCoder and Codestral maintain their lead from the baseline through iteration, suggesting that reflexion helps them make more consistent use of feedback. Overall, these results indicate that the reflexion loop effectively assists in vulnerability repair, reducing ICD-detected security issues without adding instability or fluctuations to the model outcomes.

% Figure~\ref{fig:Reflexion} graphically shows each model's results across all rounds: each line graph demonstrates \textit{Repair@}$t_i$, \textit{Regression@}$t_i$, and \textit{NetGain@}$t_i$ in percentage and the numbers shown at each $t_i$ are accuracy values. The largest improvements occur at $t_1$: for example, QwenCoder repairs nearly 8\% of insecure cases while introducing fewer than 1\% regressions, lifting accuracy from about 73\% at $t_0$ to just over 80\% at $t_1$; Codestral follows a similar trajectory. DeepSeekCoder and CodeLlama also gain at $t_1$, albeit more moderately, whereas OpenCoder shows minor improvement. Later rounds yield smaller, less consistent gains: by $t_3$, QwenCoder reaches 85\% and Codestral about 83\%, but increases from $t_2$ to $t_3$ are only a few points and regressions increase to roughly 1–2\%. DeepSeekCoder and CodeLlama plateau around 77–80\% with shrinking NetGain, and OpenCoder struggles to surpass 72\%. Overall, one or two rounds capture most of the benefit; additional iterations offer marginal improvements and occasionally undo earlier repairs.

Figure~\ref{fig:Reflexion} graphically shows each model's results across all reflexion rounds. Each line graph demonstrates \textit{Repair@}$t_i$, \textit{Regression@}$t_i$, and \textit{NetGain@}$t_i$ in percentage, and the numbers shown at each $t_i$ represent the corresponding accuracy values. These visual trends illustrate how the models gradually improve their code security through iterative reflexion and self-correction. The largest improvements occur at $t_1$: for example, QwenCoder repairs nearly 8\% of insecure cases while introducing fewer than 1\% regressions, lifting accuracy from about 73\% at $t_0$ to just over 80\% at $t_1$. Codestral follows a similar trajectory with a comparable balance between repair and regression. DeepSeekCoder and CodeLlama also show clear gains at $t_1$, though slightly smaller, while OpenCoder demonstrates only minor improvement in this stage. This pattern suggests that most models benefit strongly from the first reflexion cycle, which captures the easiest and most direct fixes from the feedback provided.  Later rounds show smaller and less consistent gains. By $t_3$, QwenCoder reaches around 85\% accuracy and Codestral about 83\%, but the increases from $t_2$ to $t_3$ are only a few points, and regressions rise slightly to around 1–2\%. DeepSeekCoder and CodeLlama gradually level off around 77–80\%, showing reduced NetGain, while OpenCoder struggles to move beyond 72\%. Overall, one or two reflexion rounds capture most of the improvement, while additional iterations provide diminishing returns and occasionally undo earlier repairs. This trend indicates that short, focused reflexion loops are sufficient to achieve meaningful security improvements without excessive computation or over-correction.

We also experimented with different prompting techniques (e.g., chain-of-thought), but they did not result in any significant improvement over the zero-shot baseline.
% We also evaluated chain-of-thought and few-shot prompting, but neither yielded a meaningful improvement in performance.

\vspace*{5pt}
\begin{mdframed}[linewidth=1pt, linecolor=gray!75!black, backgroundcolor=gray!5, roundcorner=5pt]
\textbf{Summary of Reflexion:} Reflexion increases average secure-generation accuracy from 70.74\% at $t_0$ to 79.43\% at $t_3$ (+8.69\,pp) across all models. Most gains occur at $t_1$ (+6.00\,pp); later rounds add smaller improvements (+1.66, +1.03\,pp) with slight regressions ($\approx 1\text{--}2\%$). QwenCoder and Codestral see the largest gains ($\approx +11.00\,\mathrm{pp}$), OpenCoder improves modestly, and one--two rounds capture most of benefits.
\end{mdframed}

\vspace*{3pt}
\section{Discussion and Implications} 
\label{sec:Study_Implication}

\textbf{Model Performance:} Overall, secure code generation rates tend to cluster in the low-to-mid 70\% range across models, with QwenCoder consistently outperforming others in our experimental setup. However, the level of difficulty of generating secure code varies significantly across different CWE categories and programming languages. We observe two major trends from our analysis of RQ1 and RQ2. First, a considerable proportion of first-pass generations are insecure. Even the best-performing model still produces insecure outputs for approximately one in four code completions at the initial generation point ($t_0$). Second, applying a brief reflexion loop substantially reduces this insecurity rate in the first iteration, though subsequent rounds yield diminishing returns.   
These findings mirror earlier work showing that LLM-generated code is frequently insecure by default. Our observed first-pass insecurity rate of 29\% aligns well with previously reported ranges of 25--40\%~\cite{pearce2025asleep, bhatt2023purple}. In addition, programming languages such as Python and C++ tend to yield more secure code compared to C\#~\cite{bhatt2023purple, fu2024constrained}. Finally, our breakdown by CWE type reveals a similar pattern observed by other researchers~\cite{siddiq2022securityeval, hariharan2024rethinking}. Injection-related vulnerabilities such as CWE-94, CWE-95, and CWE-798 are more easily avoided by current models. In contrast, cryptographic and arithmetic flaws, particularly CWE-338 and CWE-680, remain persistently difficult to mitigate.

\noindent 
\textbf{Reflexion:} We adopt a three-round reflexion strategy, following the approach outlined by Shi~\cite{shi2024can}. Reflexion serves as a simple, training-free mechanism that consistently improves the security of generated code, as demonstrated in our RQ2 results. The most substantial gains occur after the first reflexion round, but it does not eliminate risk.
The observed patterns across CWE types and programming languages suggest where further investment could be most impactful. For instance, improving secure defaults and curating high-quality training data are especially critical for CWE related to cryptography and memory safety types. Additionally, applying targeted reflexion for high-risk vulnerability categories, along with integrating static\cite{bhatt2023purple} and dynamic\cite{peng2025cweval} checks, can help achieve higher performance.

%\vspace*{-4pt}
\section{Threats to Validity}
\label{sec:ThreatsToValidity}
\vspace*{-3pt}

\textbf{Internal Validity:} Our results are grounded in ICD ruleset. While static analysis enables scalable, rule-based assessment, it can miss contextual cues or produce false positives. Consequently, we adopted the \emph{corrected} CyberSecEval i.e., Instruct Prime~\cite{hariharan2024rethinking}, which eliminates known compliance shortcuts and cue-based samples that could otherwise inflate performance metrics. Security outcomes can also vary depending on decoding configurations~\cite{fu2024constrained}. To ensure consistency, we fixed decoding hyperparameters across all models and reflexion rounds. Additionally, our pipeline includes an LLM-based judge to filter out non-code outputs. Misclassifications at this stage could introduce unintended bias. To mitigate this risk, we conducted manual reviews of LLM generations and adhered to validated post-processing established in prior work~\cite{tafreshipour2024prompting, pister2024promptset}. 

\noindent 
\textbf{External Validity:} Our findings are specific to \emph{instruction-style}, code-level generation, which is based on the corrected CyberSecEval–Instruct split~\cite{hariharan2024rethinking}. These results may not generalize to other settings, such as autocomplete-style generation used in the original CyberSecEval\cite{bhatt2024cyberseceval} benchmark, system-level threat assessments in CyberSecEval~2~\cite{bhatt2024cyberseceval}, or repository-scale vulnerability analyses\cite{dilgren2025secrepobench}. The findings presented in this study are limited to the eight programming languages and about 50 CWE categories included in the Instruct Prime split. While the dataset corrects for known cue contamination, the prompt distribution remains imbalanced. For instance, CWE-338 is represented by 152 samples, whereas CWE-306 and CWE-319 are represented by only one and two samples, respectively. These disparities introduce potential skew and limit the model's exposure to certain vulnerability types.

\vspace*{-3pt}
\section{Conclusions}
\label{sec:Conclusion}
% In this study, we evaluate secure code generation using a corrected benchmark that eliminates artifacts while preserving multilingual and CWE-diverse coverage. We integrate a simple reflexion loop allowing models to review and revise their own code with CWE-aligned hints. Our results show that insecurity remains prevalent: about one-quarter to one-third of first-pass outputs ($t_0$) are insecure, with notable patterns across CWE types and languages. Cryptography and input-parsing weaknesses dominate failures, whereas injection-style issues (e.g., XSS) are handled more reliably. Higher-level languages like Python tend to yield safer code than lower-level ones such as C or C\#. Reflexion consistently improves outcomes—average secure-generation accuracy rises from 70.74\% at $t_0$ to 79.43\% at $t_3$, with gains tapering after the first round. Based on Repair, Regression, and NetGain trends, we recommend running one reflexion round by default and continuing only for unresolved cases until improvements plateau.

In this study, we evaluate secure code generation on a large set of examples using a corrected benchmark that removes known artifacts and preserves multilingual, CWE-diverse coverage. We pair this setting with a simple reflexion loop that lets models review and revise their own code using CWE-aligned hints. Our findings show that insecurity remains common in first-pass generations: roughly one-quarter to one-third of outputs are insecure at $t_0$, with clear patterns by CWE and language. Cryptography and input-parsing related weaknesses account for a disproportionate share of failures, while issues with standardized, local fixes such as injection-style issues, cross-site scripting are handled more reliably. We also observed that the outcomes vary across languages: higher-level ecosystems (e.g., Python) tend to yield safer code than lower-level or configuration-heavy settings (e.g., C, C\#). Reflexion narrows this gap for all models we tested. On average, secure-generation accuracy increases from 70.74\% at $t_0$ to 79.43\% by $t_3$, with the largest gains in the first round followed by diminishing returns. From the observed patterns of Repair, Regression, and NetGain across rounds, we propose a policy as follows: run one round by default, continue only for the cases that remain insecure, and stop once NetGain shows little improvement or when regressions get close to the level of repairs.

% ---------- References ----------
\bibliographystyle{IEEEtran}
\bibliography{refs}

\end{document}